\documentstyle[aps,eqsecnum,epsf]{revtex}

\begin{document}
\setcounter{page}{0}
\renewcommand{\d}{{\rm d}}
\newcommand{\const}{{\rm const}}
\large

\title{{\rm \hfill Preprint JINR E2-98-238 \vspace{1.5cm}}\\\Large\bf
Evolution of Nonlinear Perturbations Inside\\
Einstein-Yang-Mills Black Holes
}
\author{Evgeni E. Donets\footnote{e-mail:edonets@sunhe.jinr.ru}}
\address{Laboratory of High Energies, JINR, 141980 Dubna, Russia, }

\author{Mikhael N. Tentyukov\footnote{e-mail:tentukov@thsun1.jinr.ru} 
and Mirian M. Tsulaia\footnote{e-mail:tsulaia@thsun1.jinr.ru Permanent address: \\
\hspace*{1em}Institute of Physics, Tamarashvili 6, Tbilisi 380077, Georgia}}
\address{Bogoliubov Laboratory of Theoretical Physics,
 JINR, 141980 Dubna, Russia}

\maketitle
\vspace*{2cm}

\begin{abstract}\large
    We present our results on numerical study of
    evolution of nonlinear perturbations inside spherically
    symmetric black holes in the $SU(2)$ Einstein-Yang-Mills (EYM)
    theory. Recent developments demonstrate a
    new type of the behavior of the metric for EYM black hole
    interiors; the generic metric exhibits an infinitely oscillating
    approach to the singularity, which is a spacelike but
    not of the mixmaster type. The evolution of various types of
    spherically symmetric perturbations, propagating from the
    internal vicinity of the  external horizon towards the singularity
    is investigated in a self-consistent way using an adaptive numerical
    algorithm. The obtained results give a strong numerical evidence in
    favor of nonlinear stability of the generic EYM black hole interiors.
    Alternatively, the EYM black hole interiors of S(chwarzschild)~--type,
    which form only a zero measure subset in the space of all internal
    solutions are found to be unstable and transform to the generic
    type as perturbations are developed.
  \end{abstract}

\thispagestyle{empty}
\newpage
\section{Introduction}
According to the proved singularity theorems \cite{singular}
the space-time singularities are the
most generic features of Einstein's equations. On the other hand,
the nature of the space-time singularity is model - dependent,
and still no definite answer about its most generic type exists.

 It is believed that the mixmaster type singularity   \cite{mixmaster},
which is space-like, local and oscillatory
can  pretend to be a generic one in the general cases of a gravitational
collapse and in spatially homogeneous cosmological models \cite{hom}.
The  numerical works (see \cite{isenberg}
{\em and ref's therein}) support this statement  for  some
spatially inhomogeneous cosmological models too.

 However, the recent studies \cite{dgz,dgz1,blm}
of an internal structure of spherically
symmetric non-Abelian $SU(2)$ Einstein-Yang-Mills (EYM) black holes
exhibit a new rather unexpected type of the corresponding
singularity, which is space-like, infinitely oscillating but not
of the mixmaster type. This behavior of the metric is caused by
the nonlinear nature of a source (Yang-Mills) field
in strong-field region near the black hole singularity. Thus,
the generic singularity inside non-Abelian EYM black holes
can be a possible alternative to the mixmaster one
if a nonlinear self-interacting matter field is included.

 The black hole solutions in the EYM model
are very interesting objects for several reasons \cite{vg}.
They violate the naive no-hair conjecture and exhibit a
discrete structure for an external solutions which come from
the corresponding singular boundary-valued problem, imposed in a
region between the event horizon and the spatial infinity.
Being considered dynamically, regular Bartnik-McKinnon
solitons \cite{bk} (limited cases of EYM black holes
in a limit of a shrinking event horizon) are found to be unstable
both linearly \cite{str1} and nonlinearly \cite{strzhou}.
The corresponding non-Abelian EYM black hole solutions in an
{\em external} region are also unstable under
small linear perturbations, and there exist strong evidences that
they are unstable nonlinearly \cite{zhou}.

 The goal of the present work is to study the evolution of small but
nonlinear perturbations,  arising and propagating in the internal region of
non-Abelian EYM black holes towards the singularity. We solve the full
system of self-consistent EYM evolution equations using some
kind of an adaptive mesh refinement method for
numerical simulations.

 The dynamics of small perturbations in black hole interior regions
were investigated first  for the  Reissner-Nordstr\"om
black holes \cite{rn,rn1}. The qualitative predictions
of an unbounded growth of perturbations near the Cauchy horizon
were confirmed later in a rigorous self-consistent analytical approach by
W. Israel and E. Poisson \cite{ip}.

Our investigations show that small perturbations which evolve inside
non-Abelian EYM black holes of the generic type do not grow unboundedly
and it allows us to put forward the conjecture, that unlike to EYM
black hole solutions in an external (weak field) region, the corresponding
generic (oscillating) internal solutions are stable, while an exceptional
Schwarzschild (S) - type internal solutions transform to a generic one
under the influence of nonlinear perturbations.
Thus, the generic (oscillating) type of the space-time can  be a
final stage of a spherically symmetric collapse of the Yang-Mills field
in an internal region of an acquired EYM black hole.

Recently Choptuik, Chmaj and  Bizo\'n \cite{chopt}
have studied the collapse of the self-gravitating YM field
[see also \cite{gund}]. They have investigated the external area of
the collapsing matter up to the horizon formation.
To penetrate under horizon it is necessary to use some kind of
null coordinates \cite{burko} in order to give a final
answer on the question about the nature of the resulting
space-time singularity.

 In Section II we write down the full system of EYM equations
and discuss imposed initial conditions and background
configurations; in Section III we briefly describe our numerical
algorithm; in Section IV we discuss the evolution of perturbations
inside generic (oscillating) EYM black hole interiors, and in
Section V -- inside S-type interiors; Section VI contains
the conclusions.

\section{The model and field equations}

We use a spherically symmetric purely magnetic $SU(2)$
ansatz for the Yang-Mills (YM) connection
\begin{equation}
\label{su2ansatz}
A = [ f(t,r) - 1 ](T_\phi \d\theta - T_\theta\sin \theta\d\phi ),
\end{equation}
where $T_\phi$ and $T_\theta$ are spherical projections of the
$SU(2)$ generators. The general ansatz for the spherically symmetric
$SU(2)$ YM field admits also the second independent function,
originated from the $A_0$ component of connection, but
we omit it here.

The four-dimensional spherically symmetric metric tensor also admits
two independent functions. Therefore we can
 choose the following parametrization
of the interval:
\begin{equation}
\label{ganzats}
  ds^2= \frac{\Delta\sigma^2}{r^2} dt^2  - \frac{r^2}{\Delta} dr^2 -
r^2 d\theta^2 - r^2\sin^2\theta d\phi^2.
\end{equation}
Both metric functions $\Delta$ and  $\sigma$ as well as the YM function
$f$ depend on $r$ and $t$ variables.

Let us denote
$$D =\frac{\Delta}{r},\quad\Pi=\frac{r\dot f}{D \sigma}
,\quad \phi= f ',\quad (\prime{} \equiv \partial_r,\quad
\dot{} \equiv \partial_t) .
$$

After that the full set of EYM equations looks as follows:
\begin{eqnarray}
\label{a1}
1-D ' &=& \kappa\frac{D }{r}\Pi^2
+\kappa \frac{D }{r}\phi^2+\kappa\frac{(f^2-1)^2}{2r^2},\\
\label{a2}
\frac{\sigma'}{\sigma}&=&\frac{\kappa}{r}(\phi^2+\Pi^2),\\
\label{b1}
\dot D &=&-2\kappa\frac{\sigma\Pi\phi D ^2}{r^2},\\
\label{b2}
\dot f & =& \frac{D \sigma}{r}\Pi,\\
\label{b3}
f'&=&\phi,\\
\label{c1}
\left(\frac{D \sigma}{r}\phi\right)' - \dot \Pi&=&\sigma\frac{
f(f^2-1)}{r^2},\\
\label{c2}
\left(\frac{D \sigma}{r}\Pi\right)' - \dot \phi&=&0.
\end{eqnarray}

Note, that equation (\ref{a1}) corresponds to the $tt$ component
of the Einstein equations $G^t_t = \kappa T^t_t$, equation (\ref{a2}) --
to the difference of $tt$ and $rr$ components
$G^t_t-G^r_r = \kappa(T^t_t-T^r_r)$,
equation (\ref{b1}) is the $rt$ component $G^r_t = \kappa T^r_t$ of
the Einstein equations,
equation (\ref{b2}) is the definition of $\Pi$,
equation (\ref{b3}) is the definition of $\phi$,
equation (\ref{c1}) is the Yang--Mills equation of motion, and equation
(\ref{c2}) is the requirement of $f$ to be smooth:
$ \frac{\partial}{\partial r}\frac{\partial f}{\partial t}=
\frac{\partial}{\partial t}\frac{\partial f}{\partial r}$.

It is well-known that the Schwarzschildean radial coordinate $r$
becomes the temporal coordinate, while $t$ becomes the spatial one
in the region under the event horizon of a black hole
and the dynamics in the black hole interior region is described by
the evolution equations along $r$, together with constraints at each
slice $r=\const$.

 Now (\ref{a1}), (\ref{a2}) and (\ref{b3}) are evolution
equations, (\ref{c1}) and (\ref{c2}) are wave equations,
whereas (\ref{b1}) and (\ref{b2}) occur to be conserved
constraints, which hold automatically. Indeed, after differentiating
(\ref{b2}) with respect to $r$ one has identically
$$
\left( \dot f - \frac{D \sigma}{r}\Pi\right)'=\frac{\partial}{\partial t}f'-
\left(\frac{D \sigma}{r}\Pi\right)'=\dot \phi-\dot\phi=0.
$$
Following the same lines, if we denote the relevant combination
in (\ref{b1}) as
$$
\gamma=\dot D +2\kappa\frac{\sigma\Pi\phi D ^2}{r^2},
$$
then it can be easily shown that
$$
\gamma'=\gamma\frac{\sigma'}{\sigma}.
$$
So, if $\gamma = 0$ at the initial surface $r=\const$ 
then $\gamma$ will be zero during the evolution along $r$.
As a result, there are no dynamical constraints in
our EYM system in the black hole interior region; both constraints are kept
during the evolution automatically and the system is effectively described
by the  equations
(\ref{a1}), (\ref{a2}), (\ref{b3}), (\ref{c1}) and (\ref{c2}).

However, to realize the numerical algorithm, a little different
representation of unknown functions turns out
to be more effective. Let us introduce an auxiliary field
\begin{equation}
\chi = \Delta\sigma/r  \\
\end{equation}
\noindent
as a dynamical variable instead of $\Delta$. Unlike $\Delta$,
the field $\chi$ does not exhibit oscillations  for the background
generic solutions in the interior regions of EYM black holes;
in terms of $\chi$ the local speed of light
(the slope of the characteristics) is equal to $\frac{\chi}{r}$.

Then, approaching the black hole singularity at $r \rightarrow 0$, it
is more suitable to use an inverse coordinate $R=1/r$ for the numerical
integration.

Now the resulting set of unknown functions, used for numerical study
of our PDE system consists of
$$
\chi=\Delta\sigma R,\quad
\sigma,\quad
\phi = \chi R^3 f',\quad
\Pi = \dot f
$$
(here and below $\prime{} \equiv \partial_R$ ),
and the full system of equations obtained from (\ref{a1}), (\ref{a2}),
(\ref{c1}), (\ref{c2}) finally has the form:
\begin{eqnarray}
\chi'&=& \sigma\left(\frac{\kappa}{2}(f^2-1)^2 - \frac{1}{R^2}\right),\label{eq1}\\
\frac{\sigma'}{\sigma}& = & - \frac{\kappa}{R^3\chi^2}(\phi^2+\Pi^2),\label{eq2}\\
\phi'-\frac{\partial}{\partial t}\left(\frac{\Pi}{R^3\chi}\right)
&=&\sigma f(f^2-1),\label{eq3}\\
\Pi'-\frac{\partial}{\partial t}\left(\frac{\phi}{R^3\chi}\right)&=&0,
                                                                 \label{eq4}\\
f'&=&\frac{\phi}{R^3\chi}.\label{eq5}
\end{eqnarray}

Apart from these equations we also use the constraints
\begin{eqnarray}
\Pi&=&\dot f , \label{constr1}\\
\frac{\dot \sigma}{\sigma}&=&\frac{\dot \chi}{\chi}+2\kappa\frac{\Pi\phi}{\chi}
  \label{constr2}
\end{eqnarray}
in order to set the initial data and to keep the control of the accuracy.

 We set the coupling constant $\kappa=1$ hereafter without loss
of generality.

 After the completion of numerical calculations we display
the results again in terms of metric function $\Delta(r,t)$ to get
more insights on their physical meaning and to compare them with the
corresponding background configurations.

\subsection{\large \bf Background configurations}

We study the evolution of perturbations, propagating
on a homogeneous ($t$~-independent) background configurations  which
correspond to the interiors of spherically symmetric EYM black
holes. These configurations are the solutions of the system
\begin{eqnarray}
\nonumber \Delta(f'/r)'+\left(1-\frac{(f^2-1)^2}{r^2}\right)f'&=&
            \frac{f(f^2-1)}{r}\\
\label{24PRD}\left(\frac{\Delta}{r}\right)'+
              2\Delta\left(\frac{f'}{r}\right)^2&=&1-\frac{(f^2-1)^2}{r^2}\\
\nonumber (\ln \sigma)'&=&\frac{2{f'}^2}{r}
\end{eqnarray}
obtained from (\ref{c1}), (\ref{a1}) and (\ref{a2}), neglecting the
$t$~-dependence in the domain $r<r_h$. The initial conditions are imposed
in the small vicinity of the surface $r=r_h$ which is implied to be the
position of a simple (not double) event horizon. As it was shown recently
\cite{dgz,dgz1,blm}, the generic solution of the system (\ref{24PRD}),
corresponding to the interior of a static spherically symmetric EYM
black hole, has no Cauchy horizons, and the metric exhibits
an infinitely oscillating behavior (but not of the mixmaster type)
with an amplitude, unboundedly growing towards the space-like
singularity.

The oscillating structure
of the metric for the generic solution originates from the
features of the corresponding 2-dimensional autonomous
dynamical system. This system effectively describes generic solutions
in the regime when some irrelevant terms in  (\ref{24PRD})
($1$ is negligible in comparison to $(f^2-1)^2/r^2$ and the YM function $f$
is set equal to constant, $f=\const$, while $\partial_r f \ne 0$ remains
dynamical ) are neglected near $r \rightarrow 0$ \cite{dgz}.
In the interior of the EYM black hole the metric passes through
an infinite series of ``almost Cauchy horizons'' in the maxima of
oscillations which alternate by subsequent huge falls of the metric
function $\Delta$ in the minima;
the frequency of oscillations of the metric
exponentially grows as the singularity is  approached.
The approximate recurrence formulas, obtained in  \cite{dgz1},
allow one to describe the behavior of the EYM system in such a ``strong
oscillation'' regime with an accuracy, improving towards the
singularity.

However, for a typical generic EYM internal black hole
solution the  ``strong oscillation''
regime  described above
does not start just in a vicinity of the event horizon.
Depending on the initial conditions on the event horizon,
before the first huge fall of the ``strong oscillation'' 
regime begins at some
$r=r_{osc.}$, the solution is determined by the complete system
(\ref{24PRD}) with all terms relevant in the significant
domain $r_{osc.}<r<r_h$. In this domain the metric function $\Delta$,
being negatively defined, also can admit a few oscillations with
a small amplitude (in comparison with ``strong oscillations'');
we call this regime a ``weak oscillation'' regime.

 In the present paper we consider the evolution of perturbations
starting from an internal vicinity of the event horizon and then
propagating through the ``weak oscillation'' region and the first
huge fall
of the metric function $\Delta$ in the ``strong oscillation'' regime
up to the next ``almost Cauchy horizon''.

 As it was also shown in \cite{dgz,blm}, for some discrete
values of initial parameters, EYM spherically symmetric
black holes also admit the standard Schwarzschild 
and Reissner-Nordstr\"om (RN) interiors
(third, so called HMI internal solution is not
asymptotically flat and we do not consider it here);
however such interiors are rather exceptional cases and
they form only a subset of a zero measure
in the space of all EYM internal black hole solutions.
The evolution of perturbations inside S~-type EYM black holes
is  investigated as well and the results will be discussed below.

\subsection{\large \bf Initial conditions}

The Cauchy problem for the system of equations (\ref{eq1})-(\ref{eq5})
can be set as follows:
we set $f,\quad\phi,\quad\chi\,$ at some space-like surface $r=\const$ and
define $\Pi$ and $\sigma$ from (\ref{constr1}), (\ref{constr2}).
Once being imposed, the constraints (\ref{constr1}) and (\ref{constr2}) will be
satisfied during the evolution along $r$ due to the equations
(\ref{eq1}) - (\ref{eq5}). The latter can be solved using a standard
finite - difference numerical technique.

Note, that we suggest the initial space-like hypersurface $r=\const<r_h$
 situated with the necessity under the external event horizon.
Therefore the perturbations  given at this hypersurface have
a status of fluctuations, acquiring in an internal region of a
black hole and they are not related anyhow with those,
propagating inward from an external region through an event
horizon.

The initial values of $f$ and $\phi$ can be set independently.
Straumann and Zhou \cite{strzhou} considered two types of initial
perturbations for the case of Bartnik-McKinnon regular solitons
and the external region of EYM black holes. They called the perturbations
class I if only the YM function $f$ is initially perturbed,
while the time derivative of the perturbation equals to zero
$\delta\phi = 0$; it means that the initial kinetic energy of the
YM field  vanishes. In class II perturbations the function $f$ remains
unperturbed, but $\delta\phi$ and thus the kinetic energy does not
vanish initially.

 We  used both these classes of initial perturbations for
our problem in the internal region of EYM black holes. The considered
initial data come from the asympthotics near the external event horizon.
Since the horizon is {\em not} a regular point, we cannot
set all Cauchy data independently. The data must satisfy the series
expansions originated from the Einstein-Yang-Mills equations.

The first order of asympthotics for background $t~$-independent
EYM equations give
$
\Delta=d_1(r-r_h),\quad \sigma=\sigma_0+\sigma_1(r-r_h),
\quad f=f_0+f_1(r-r_h),
$
where $r_h$ is the radius of the event horizon and $f_0$, $\sigma_0$
are free parameters. Other parameters are expressed in terms of
these free parameters as follows:
\begin{equation}\label{coeffs}
d_1=r_h-(f_0^2-1)^2/r_h,\quad  f_1=f_0(f_0^2-1)/d_1,\quad
\sigma_1=2 \sigma_0 f_1^2/r_h.
\end{equation}
\noindent
Thus, in order to define the background solution, we should set the values
$r_h,\quad f_0,\quad \sigma_0\,$; the coefficients $d_1,\quad f_1\,$
and $\sigma_1$ are determined from (\ref{coeffs}).

We choose an initial surface at the distance $h$ (one $r$~- step) 
under the horizon.
The equilibrium (background) values of $f$ and $\chi$ at this surface are
$
f =f_0-f_1 h,\quad
\chi=-d_1 \sigma_0 h/r_h.
$

 {\it Class I perturbations.} In the case when
 small, for example, Gauss - like
perturbation
\begin{equation}\label{perturbations}
K e^{-s(t-t_0)^2}
\end{equation}
is added to the equilibrium YM function $f$, then
the field $\Pi$ is  determined  by the equation (\ref{constr1}),
i.e.,
$$
\Pi(t)|_{r=r_h-h}=-2 K s (t-t_0) e^{-s(t-t_0)^2}.
$$
For initial $\chi$ we use the equilibrium value and
initial $\phi$ is also set to be unperturbed. Thus, class I
initial perturbations are defined as follows:
\begin{eqnarray}\nonumber
f(t)|_{r=r_h-h}&=&f_0-f_1 h+K e^{-s(t-t_0)^2},\\ \label{classI}
\chi(t)|_{r=r_h-h}&=&-d_1 \sigma_0 h/r_h,\\
\phi(t)|_{r=r_h-h}&=&-d_1 \sigma_0 f_1 h/r_h^2
\nonumber
\end{eqnarray}
\noindent
with $\Pi(t)|_{r=r_h-h}$ determined above.

{\it Class II perturbations}. In this case the YM function $f$ remains
initially unperturbed along with the corresponding function $\Pi$,
while the function $\phi$ gets independent deviation from the background:

\begin{equation}
\begin{array}{rcl}
f(t)|_{r=r_h-h}&=&f_0-f_1 h,\\ 
\Pi(t)|_{r=r_h-h}&=&0, \\
\chi(t)|_{r=r_h-h}&=&-d_1 \sigma_0 h/r_h,\\
\phi(t)|_{r=r_h-h}&=&-d_1 \sigma_0 f_1 h/r_h^2+K e^{-s(t-t_0)^2}.
\end{array}
\label{classII}
\end{equation}

 Both classes of initial perturbations give rise to two
scattering waves. One of them propagates with time $r$ in a
positive spatial $t$~-direction (``ingoing'' wave) and another one
propagates in the negative $t$~-direction (``outgoing'' wave). After some
interference in a region around $t=t_0$ they move independently from
each other in opposite spatial directions. We investigated the evolution
of class I and  class II initial perturbations on various EYM black hole
interior backgrounds and found their behavior to be the same
after the ``ingoing'' and ``outgoing'' waves are completely divided in
the space.
Moreover, if we set the initial perturbation with a symmetry with respect
to the center of perturbation $t_0$, this symmetry is conserved during the
evolution.

 That is why one can consider without loss of generality the third class
of initial perturbations in addition to the described above;
the initial perturbations of this class determine only one initial wave,
propagating, say, in a positive $t$~-direction. This class occurs to be
a combination of class I and class II and is defined by the following way.

{\it One-wave class of initial perturbations}.
Perturbations of this class are the same as in class I for the
functions $f$ and $\Pi$. In  addition, the perturbation of $\phi$ is
chosen not independently, but in such a way, that
 the initial ``outgoing'' wave is canceled:
\begin{equation}
\begin{array}{rcl}
f(t)|_{r=r_h-h}&=&f_0-f_1 h+K e^{-s(t-t_0)^2},\\
\Pi(t)|_{r=r_h-h}&=&-2 K s (t-t_0) e^{-s(t-t_0)^2},\\
\chi(t)|_{r=r_h-h}&=&-d_1 \sigma_0 h/r_h,\\
\phi(t)|_{r=r_h-h}&=&-d_1 \sigma_0 f_1 h/r_h^2-2 K s (t-t_0) e^{-s(t-t_0)^2}.
\end{array}
 \label{pert}
\end{equation}
\noindent
We will use  this class of initial perturbations for the illustration
of their nonlinear evolution in subsequent Sections.

 Now the initial value for the field $\sigma$ for all considered classes of
perturbations can be determined by the numerical integration of
the equation (\ref{constr2}). We integrated it from the left to the right,
since the unperturbed functions are situated at the left side at $t < < t_0$.

\section{Numerical method}

The specific feature of the strong - field dynamics requires
a numerical code
which must be able to resolve a solution on extremely small scales.
Usually some kind of an adaptive mesh refinement algorithm \cite{berger}
is implemented for these problems.

We controlled the resetting of grids ``manually''. The accuracy was
determined by means of calculation of the constraint (\ref{constr2}).
When the absolute value of this constraint at least at one point
became large than $10^{-5}$ the program saved its state and stopped.
Then we investigated output data visually, determined the ``bad'' region
and set new grid steps.
Starting again, the program picked out necessary region and
recalculated the data according to new grid steps.
In order to set  $\Pi$ and $\sigma$ according to new data,
the constraints (\ref{constr1}), (\ref{constr2}) were also
recalculated.

\subsection{\large \bf Algorithm}

The system of equations (\ref{eq1}) - (\ref{eq5}) was solved by means of
MacCormac predictor - corrector scheme (see, for example,  \cite{strzhou}).
We  tested many  modifications of this scheme and convinced ourselves
that in this case
various finite differences used for forward/backward difference
lead approximately to the same accuracy.
The results can vary from one domain to another, but not very strongly.
The most universal approach appears to be the conventional
MacCormac forward/backward difference, so we used it for our
calculations.

The constraint (\ref{constr1}) does not contain derivatives of unknown
function, so we integrated it using a simple midpoint formula:
$$
\Pi_i=\frac{f_{i+1}-f_{i-1}}{t_{i+1}-t_{i-1}}.
$$

For integration of the constraint (\ref{constr2})
we  tried to use several methods, such as various Runge-Kutta schemes,
predictor-corrector methods, etc.
The best result was obtained by the simplest midpoint formula:
\begin{equation}\label{fd_constr2}
\sigma_{i+1}=\sigma_{i-1}+(t_{i+1}-t_{i-1})\frac{\sigma_i}{\chi_i}
\left(\frac{\chi_{i+1}-\chi_{i-1}}{t_{i+1}-t_{i-1}}+
4\Pi_i\phi_i\right).
\end{equation}
This result is obvious because we used centered differencing
for monitoring of the constraint (\ref{constr2}).

We calculated the maximal slope of the characteristics $\chi R^3$
on each slice $R=1/r=\const$
 and then chose $h_R$ (``time'' $R$~- step)
 according to the Courant criterion, $h_R=h_t C/2$.
We denote the ``spatial'' current integration step by $h_t$  and
the maximum of the slope of characteristics on a current slice as
$C = {\rm max}(\chi R^3)$.

The value of $C$ tends to $\infty$ as $r\to 0$ (or $R \to \infty$).
In order to prevent a destruction of the numerical algorithm, we bounded
it by some value $C_{max}$. Surely, the Courant criterion was satisfied
with guarantee for all $C_{max} < C$.
The final expression for $h_R$ looks as follows:
$$
h_R =
\left\{
\begin{array}{cl}
\frac{C \,h_t}{2},&\quad \mbox{if}\quad C<C_{max}\\
\frac{C_{max} h_t}{2},&\quad \mbox{if}\quad C \geq C_{max}
\end{array}
\right.
$$
The value of $C_{max}$ was set ``by hand'' and varied from $10$ to
$10^4$ in different regions.

\subsection{\large \bf Realization remarks}

The algorithm was realized as ANSI  {\em C} program.
For real numbers the type \verb|double| was used.

The initial ``spatial'' ($t$) step was set  equal to
 $10^{-3}$, the relevant initial
interval was $ -30.0 \leq t \leq 30.0$.
So, the number of one slice points we started with was about 6000.
During the calculation, this number may increase up to $10^6$.
In output the program saved only the part of the points,
and the distribution function of the saved points was defined
according to the estimation of an error, produced by the usage of the coarse
grid.

The initial data were calculated in the coordinate $r$ and then
converted into the coordinate $R$.

The typical time needed for evaluation
of one configuration was about 50 hours in Pentium-166.

\section{Generic solutions}

We have tested various background generic internal configurations
with values of $r_h$ in a range from $10^{-1}$ up to $10^5$,
corresponding to both asymptotically flat (black hole) and
not asymptotically flat EYM solutions in an external region.
Similarly to the background configurations, the evolution of perturbations
under the event horizon turns out not to be  sensitive to the background
type in the external region (either asymptotically flat or not) and
looks qualitatively the same for both considered types.

\begin{figure}[p]
\centerline{\vbox{\epsfysize=35mm \epsfbox{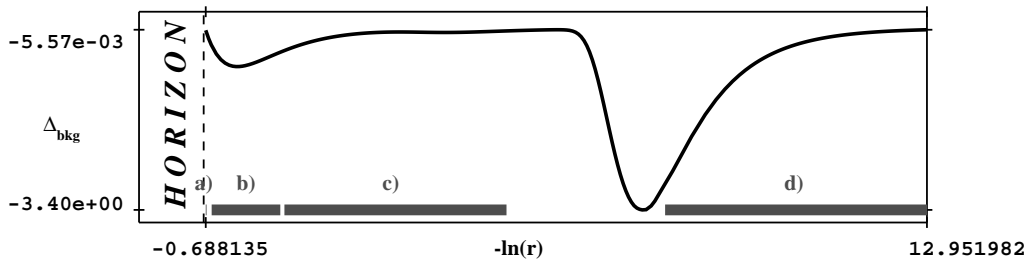}}}
\caption{\large \label{dstatic} Background $\Delta$ under the event horizon.
One can see two oscillations of the function
$\Delta$ in the ``weak oscillation'' region (second minimum at
$-\ln{r}=3.81671$ is non-observable in this scale)
and the first oscillation in the ``strong oscillation'' region.
Although the first fall of $\Delta_{background}$ is not  very huge,
it is described, however, by the reduced system of equations with a good
accuracy and therefore corresponds to the ``strong oscillation'' regime.
Thick grey lines below mark regions depicted in 3-D plots of
$\delta \Delta$: a) corresponds to Fig.\protect\ref{deltaa},
this region is very small; b) corresponds to Fig.\protect\ref{deltab};
c) corresponds to Fig.\protect\ref{deltac} and
d) corresponds to Fig.\protect\ref{deltad}.}
\end{figure}

 We were able to follow the evolution of perturbations of
both classes (I and II) with different (small enough) initial amplitudes,
propagating through the ``weak oscillation'' region
and the first huge fall of the function $\Delta$ up to a vicinity of
the first ``almost Cauchy horizon'' of the ``strong oscillation'' region.
The common result is that during propagation towards the singularity
the considered perturbations do not grow up unboundedly and
look like a light ripple on the background solutions. Therefore
the strong numerical evidence in favor of nonlinear stability of
generic (oscillation type) internal EYM black hole solutions is
obtained.

To illustrate this statement we have chosen an appropriate background
configuration. Being  asymptotically not flat in exterior,
it exhibits all important features of EYM black hole interiors and
allows one to plot both ``weak'' and ``strong oscillation''
regions in the same figure with enough resolution.
The chosen background is characterized by the following parameters
$$ \begin{array}{l}
r_h=2.0\quad (R=0.5,\quad -\ln{r}=-0.69314718);\\
f_0=-0.302072;\\
\sigma_0=0.9,
\end{array}$$
and the corresponding curve of the metric function $\Delta$ is plotted
in Fig. \ref{dstatic}.

 For the illustration purposes we have used the one-wave class of initial
perturbations. According to Section II, it is determined by a $t$~-dependent
deviation of the YM function $f$ from its background value which induces the
perturbation of $\Pi$; the perturbation of the function $\phi$ is defined
to cancel the initial wave, propagating to the negative $t$~-direction.

We plot below the evolution of perturbations with initial
parameters in (\ref{pert}):
$$
s = 100.0; \qquad t_0=-10.0
$$
for two different amplitudes $K_1=10^{-4}$ and $K_2=10^{-3}$ in
(\ref{perturbations}) (we will call them ``small'' and ``big''
perturbations respectively in the further discussion).
These perturbations induce deviations of all other functions
according to the full set of
EYM equations and the resulting perturbation looks like a nonlinear
wave, propagating in the positive spatial $t$~-direction with time $r$,
directed to $r \rightarrow 0$.

\begin{figure}[p]
\centerline{\vbox{\epsfysize=80mm \epsfbox{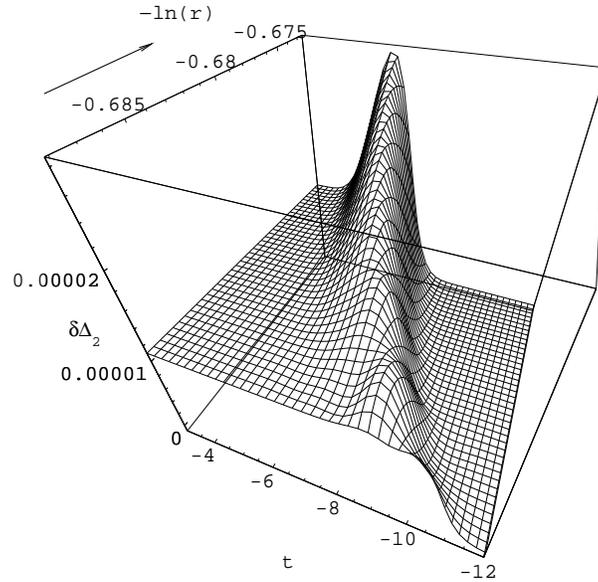}}}
\caption{\large \label{deltaa} The beginning of the evolution of perturbation of
$\delta\Delta_2 = \Delta_2 - \Delta_{background}$.
The initial perturbation looks like a step. The folds of this step
remain to be of a constant height during the beginning of evolution.
Further the left fold (greater $t$) will change while the right is
constant by definition. This figure corresponds to region a) in
Fig.\protect\ref{dstatic}. The arrow points to the direction
of the evolution.}
\end{figure}

\begin{figure}[p]
\centerline{\vbox{\epsfysize=80mm \epsfbox{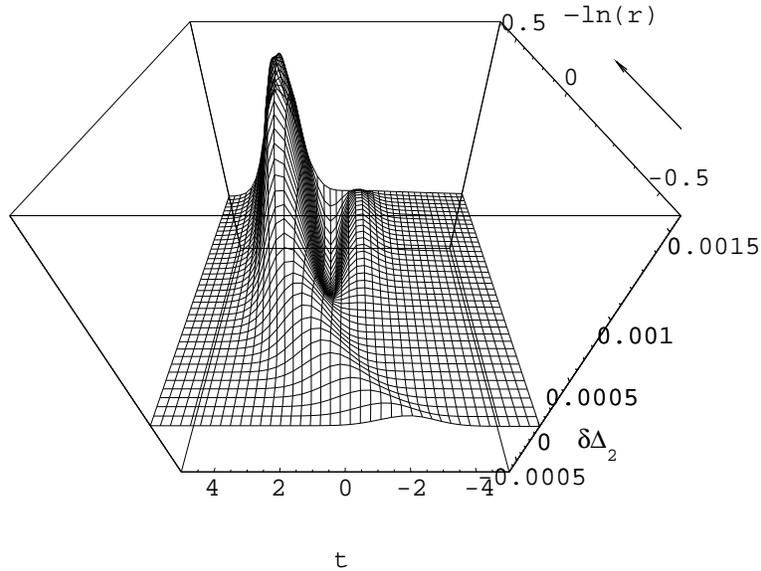}}}
\caption{\large \label{deltab} Plot of $\delta\Delta_2$ in the beginning
of the ``weak oscillation'' region. This figure corresponds to b) region
in Fig.\protect\ref{dstatic}. The arrow points to the direction of
the evolution.}
\end{figure}

\begin{figure}[p]
\centerline{\vbox{\epsfysize=80mm \epsfbox{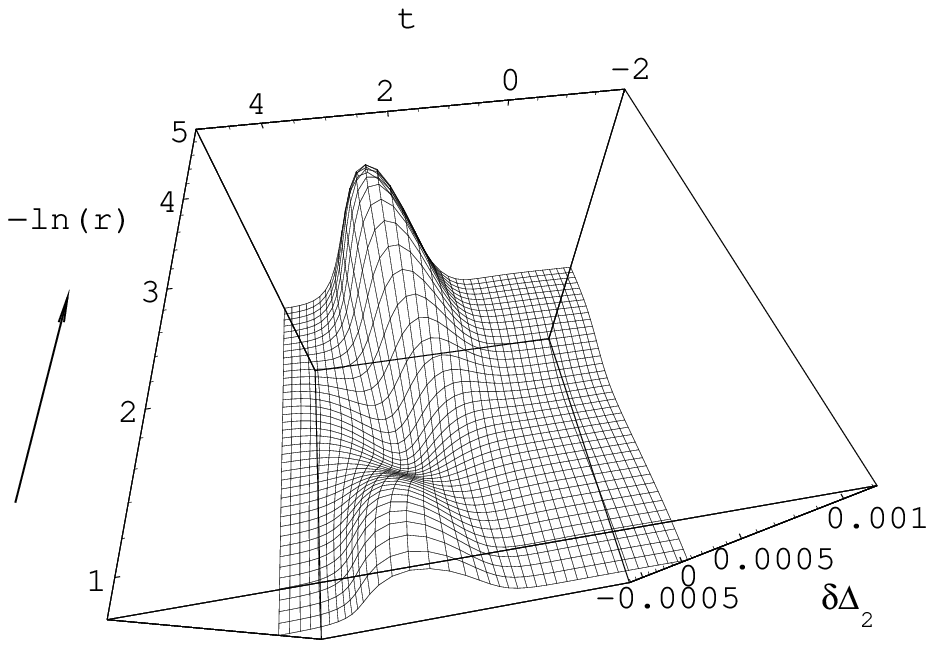}}}
\caption{\large \label{deltac} Plot of $\delta\Delta_2$ in the
``weak oscillation'' region. This figure corresponds to c) region
in Fig.\protect\ref{dstatic}. The arrow points to the direction of
the evolution.}
\end{figure}

\begin{figure}[p]
\centerline{\vbox{\epsfysize=80mm \epsfbox{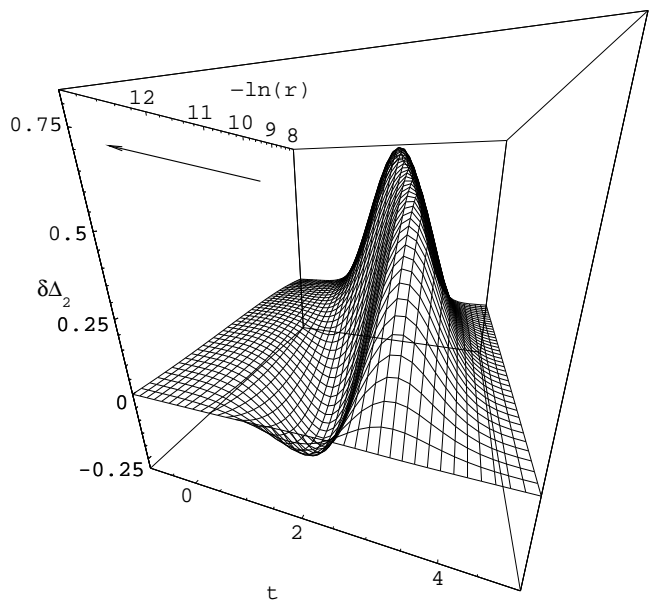}}}
\caption{\large \label{deltad} Plot of $\delta\Delta_2$ in the
``strong oscillation'' region. This figure corresponds to d) region in
Fig.\protect\ref{dstatic}. The arrow points to the direction of
the evolution.}
\end{figure}

To show the relative size and shape of perturbations one can
normalize the corresponding functions by their equilibrium values.
The value $\delta \Delta_2$, obtained as
$\delta \Delta_2 = \Delta_2 - \Delta_{background}$ for ``big''
initial perturbation, is plotted in four figures: Fig.\ref{deltaa},
Fig.\ref{deltab}, Fig.\ref{deltac} and Fig.\ref{deltad}.
The considered $r$~- regions are shown in Fig.\ref{dstatic} by thick grey
lines. The plots illustrate a strongly nonlinear nature of the evolution:
the perturbation of the metric function $\Delta$ changes the relative sign
and the shape as singularity $r \rightarrow 0$ is approached. The
perturbations of other relevant functions behave in a similar way
and we do not plot them here.

\begin{figure}[p]
\centerline{\vbox{\epsfysize=90mm \epsfbox{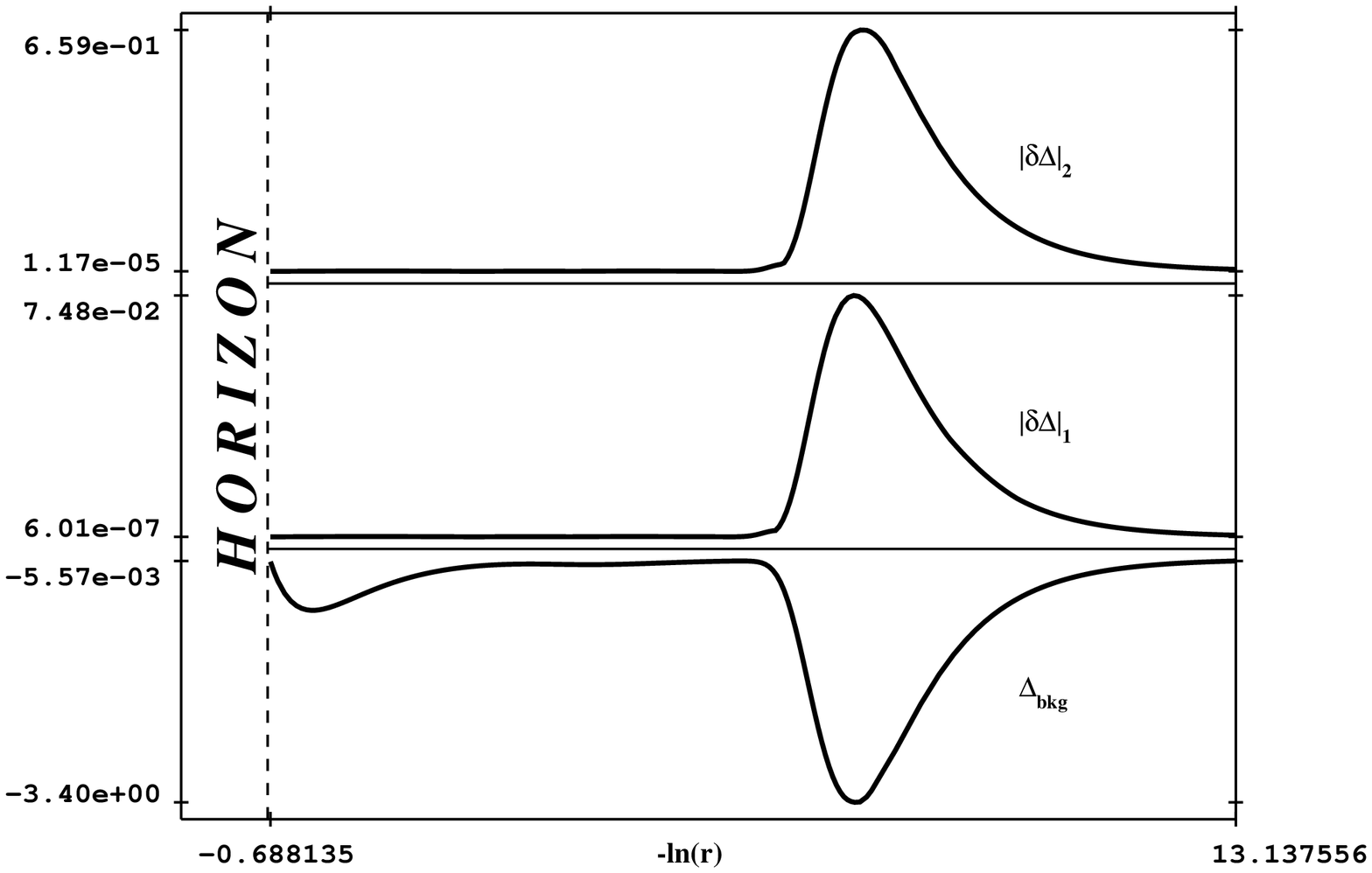}}}
\caption{\large \label{delta} $\Delta$ background (bottom), absolute value of
the maximal deviation from the background (middle -- ``small'' initial
perturbation, top -- ``big'' initial perturbation).}
\end{figure}

\begin{figure}[p]
\centerline{\vbox{\epsfysize=90mm \epsfbox{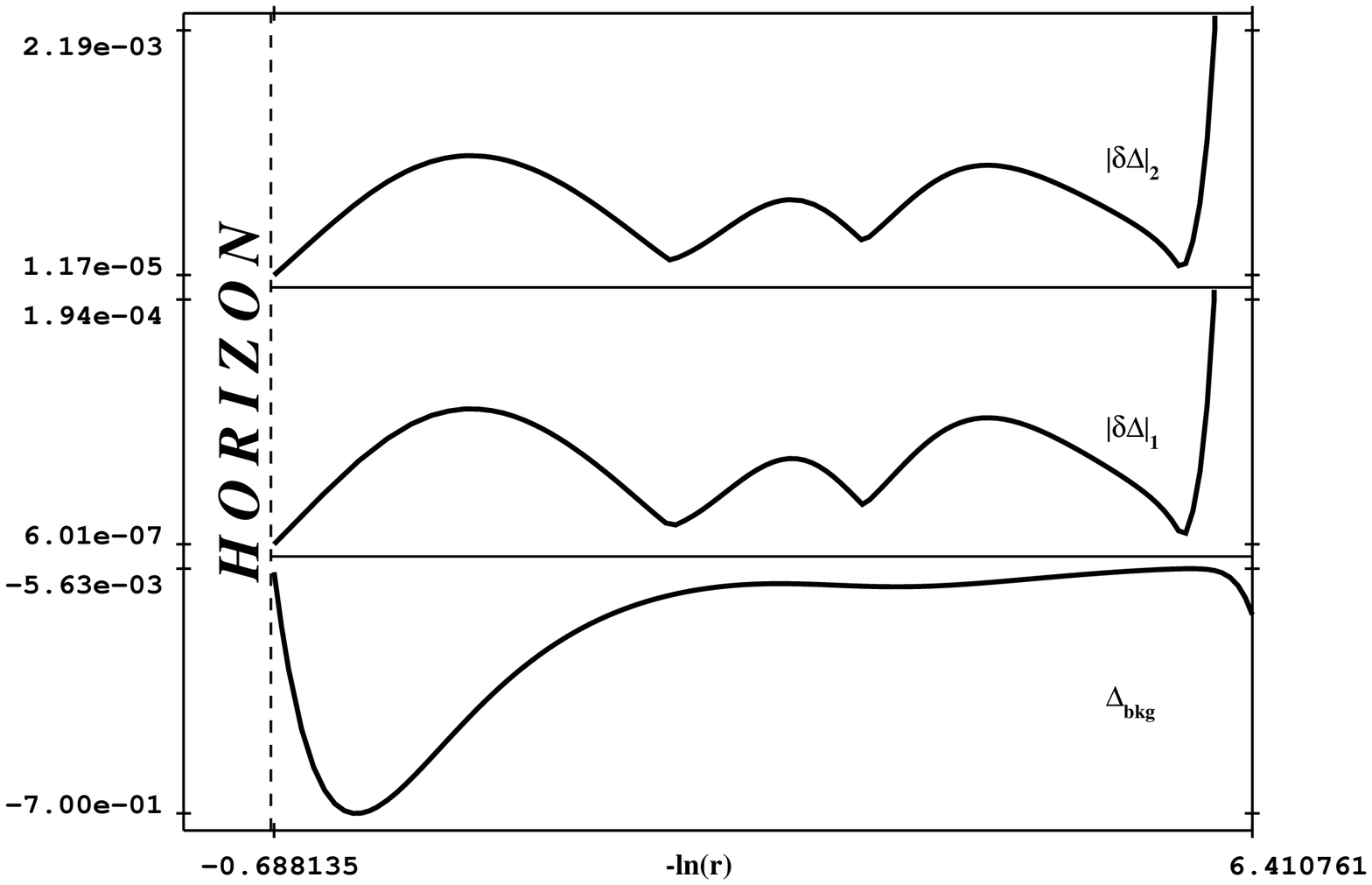}}}
\caption{\large \label{deltapart}$\Delta$ in the ``weak'' oscillation region:
background (bottom), absolute value of the maximal deviation from the
background (middle -- ``small'' initial perturbation, top -- ``big''
initial perturbation).}
\end{figure}

\begin{figure}[p]
\centerline{\vbox{\epsfysize=90mm \epsfbox{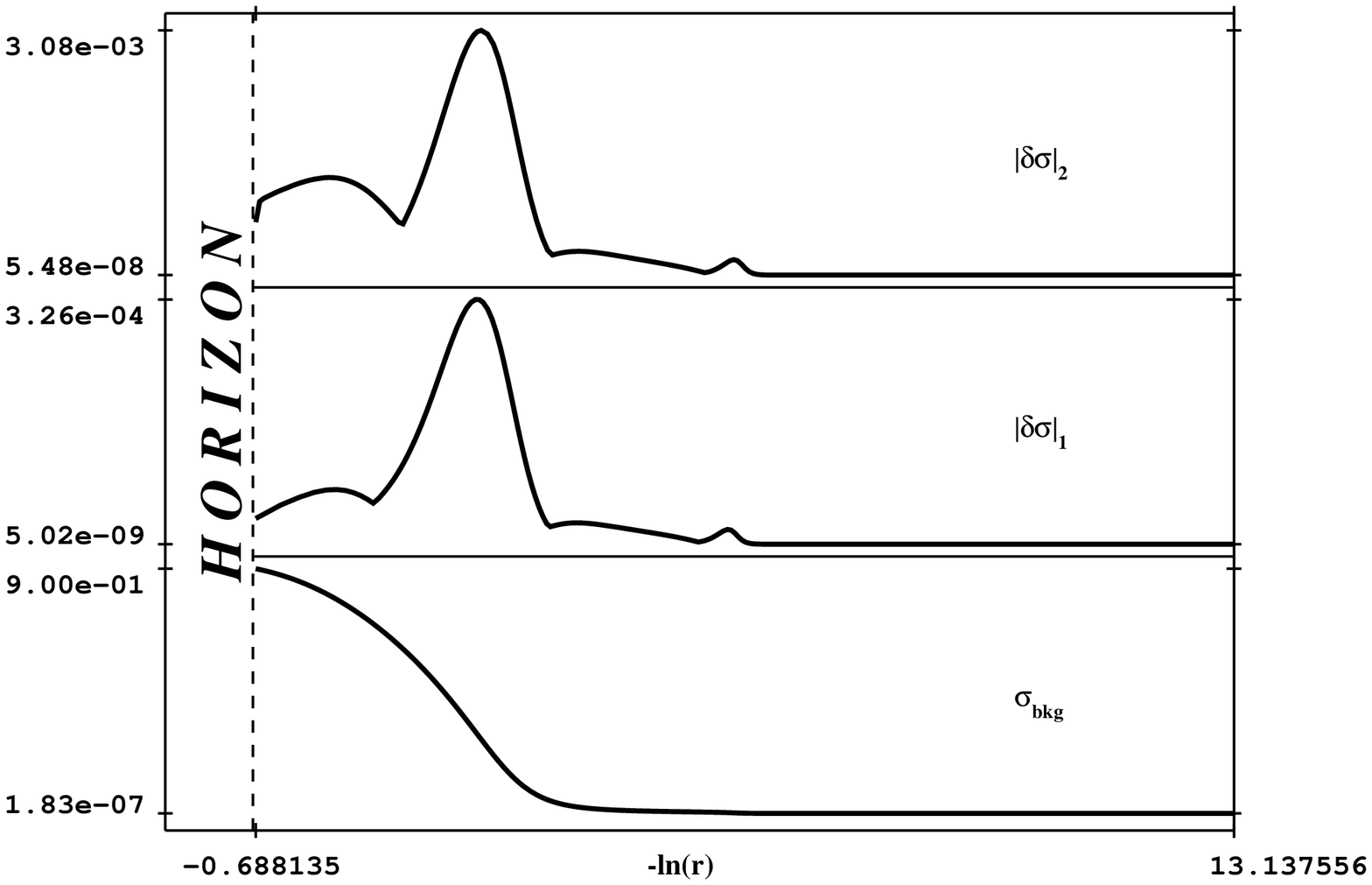}}}
\caption{\large \label{sigmafig}$\sigma$: background (bottom), absolute value of
the maximal deviation from the background (middle -- ``small'' initial
perturbation, top -- ``big'' initial perturbation).}
\end{figure}

\begin{figure}[p]
\centerline{\vbox{\epsfysize=90mm \epsfbox{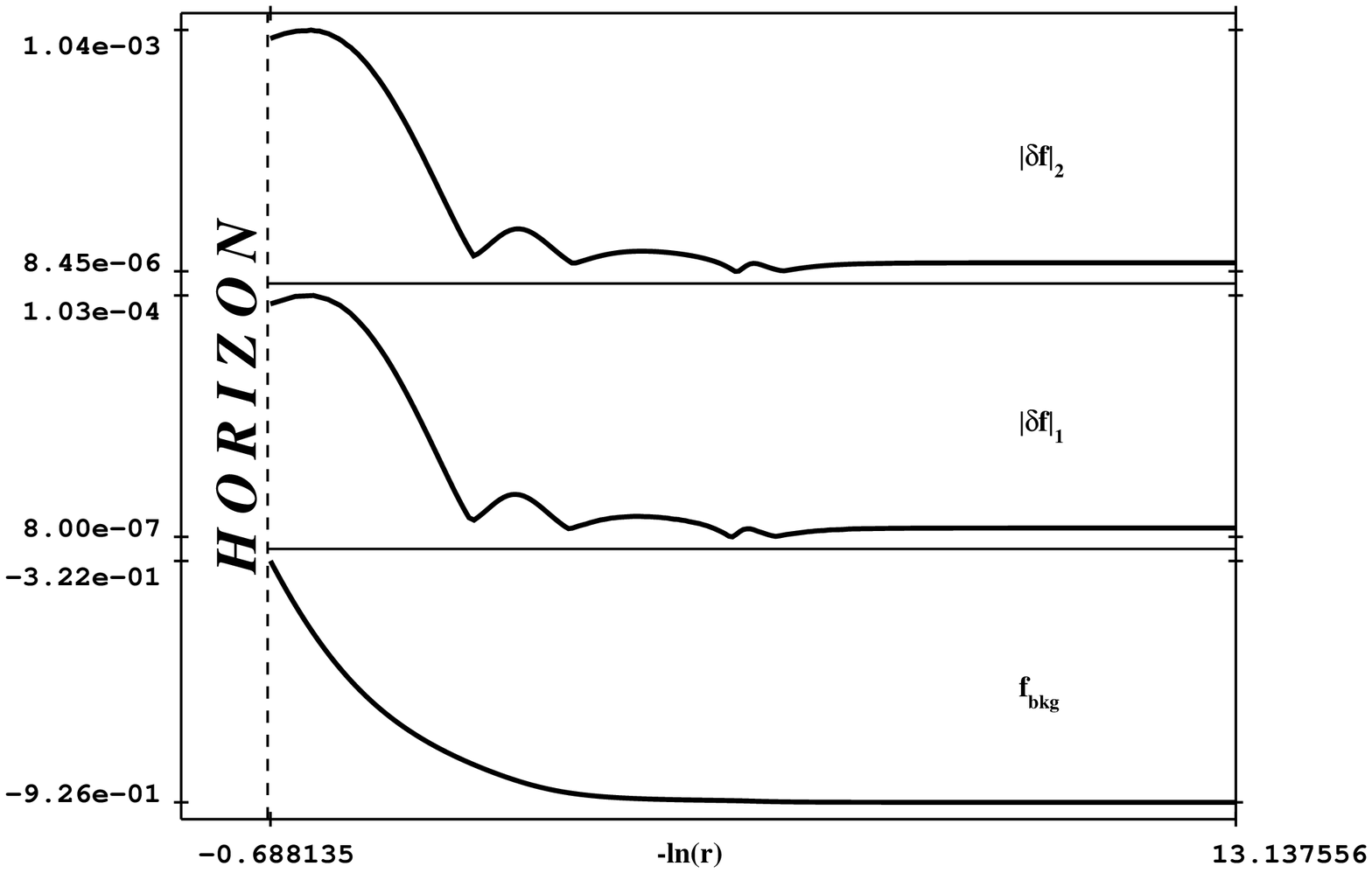}}}
\caption{\large \label{ffig}$f$: background (bottom), absolute value of
the maximal deviation from the background (middle -- ``small'' initial
perturbation, top -- ``big'' initial perturbation).}
\end{figure}

\begin{figure}[p]
\centerline{\vbox{\epsfysize=90mm \epsfbox{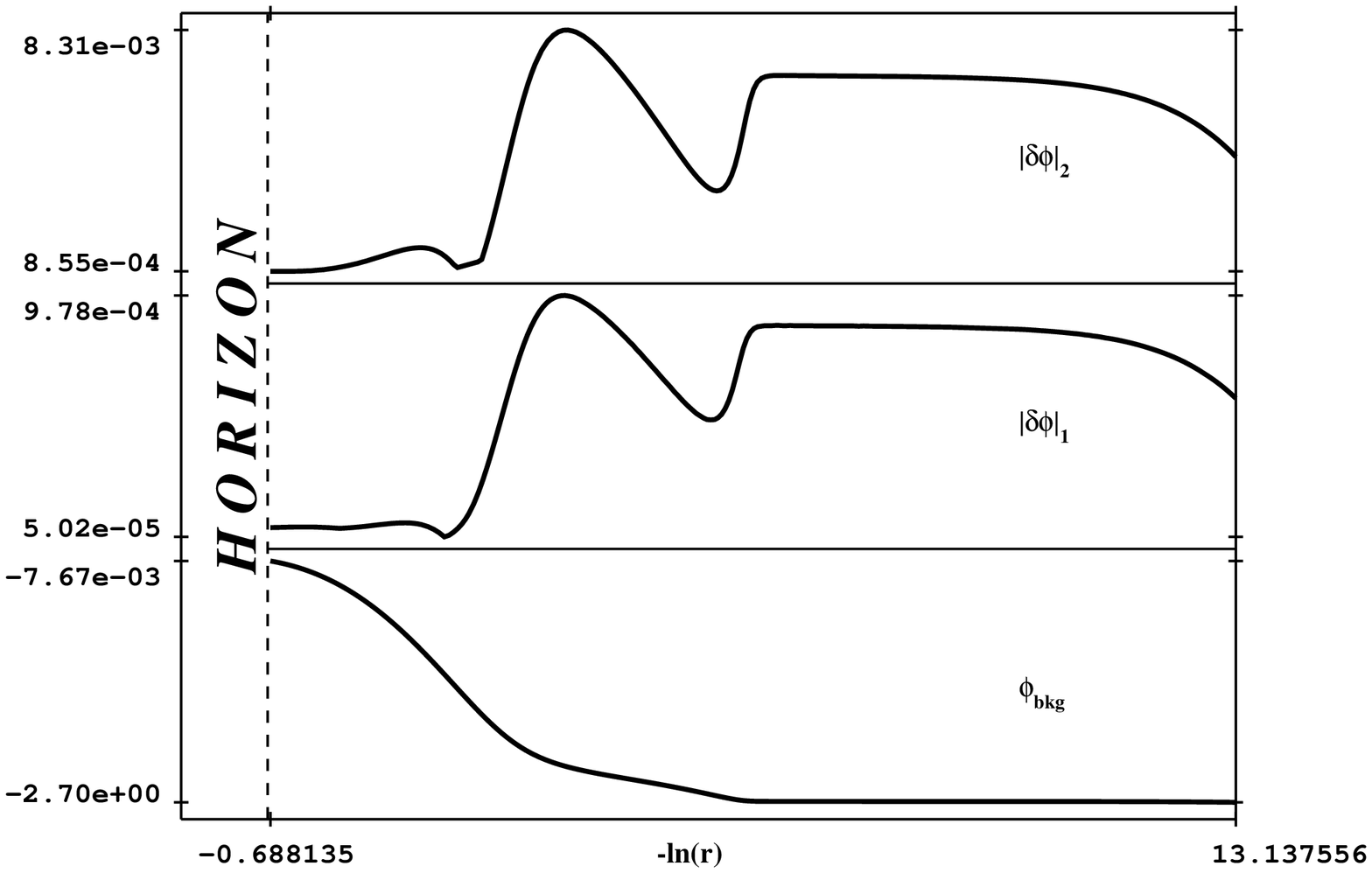}}}
\caption{\large \label{phifig}$\phi$: background (bottom), absolute value of
the maximal deviation from the background (middle -- ``small'' initial
perturbation, top -- ``big'' initial perturbation).}
\end{figure}

\begin{figure}[p]
\centerline{\vbox{\epsfysize=60mm \epsfbox{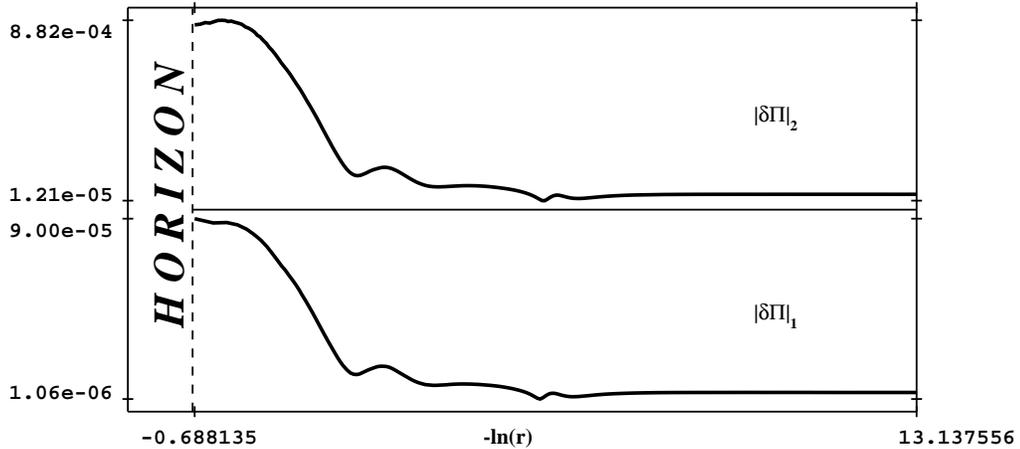}}}
\caption{\large \label{pifig}$\Pi$: absolute value of the
maximal deviation from the background (bottom -- ``small'' initial
perturbation, top -- ``big'' initial perturbation); the background value
of $\Pi$ is zero by definition.}
\end{figure}

 However, the perturbations do not grow unboundedly and therefore
they do not
destroy the background solution. It is convenient to plot
the absolute values of the maximal deviation of all relevant functions
from their backgrounds as perturbations are developed.
The corresponding plots are represented in Figures \ref{delta} -- \ref{pifig}
for both ``small'' ($K=10^{-4}$) and ``big'' ($K=10^{-3}$) initial
perturbations along with the background functions.

 These plots exhibit most general features of the perturbations,
propagating in the different considered generic backgrounds:
the evolution of ``small'' and ``big'' perturbations
looks very similar for all relevant functions, so the difference
in initial amplitude does not produce differences in shapes and
characters of perturbations; the amplitude of perturbations
of $\Delta$ is approximately proportional 
\noindent
to the absolute value of
$\Delta$ itself (it grows in the ``strong oscillation'' regime following
the fall of $\Delta_{background}$  and then decreases as
$\Delta_{background}$  approaches the ``almost Cauchy horizon'');
perturbations of all other functions demonstrate  bounded nonlinear
oscillations in the ``weak oscillation'' region, then they become almost
constant ones in the ``strong oscillation'' region up to a vicinity
of an ``almost Cauchy horizon''.

 We have followed the evolution of the perturbations towards 
the $r=0$ singularity up to the first ``almost Cauchy horizon''
in the ``strong oscillation'' regime. 
%
However,
there are no physical reasons to consider the next huge oscillation
of the metric function $\Delta$, since in this region the magnitude of the Riemannian squared
scalar exceeds the Planckian value ($R_{\mu \nu \lambda \tau}R^{\mu \nu \lambda \tau} > 1/L^4_{pl.}$)
and the classical description of space-time is no longer valid overthere.
So, the only problem remains to penetrate numerically through the first
``almost Cauchy horizon'' in the ``strong oscillation'' regime.
We are going to consider this problem separately along with RN~-type 
using the more precise numerical code.

\section{Schwarzschild -- type solutions}

Schwarzschild and Reissner-Nordstr\"om types of internal
background EYM black hole solutions are of exclusively types
since they form a set of zero measure in the space of all initial data
\cite{dgz},\cite{blm}. They exist only for some discrete values of
 ($r_h$, $f_0$)
on a horizon and any internal background solution of (\ref{24PRD})
with an arbitrary small deviated initial data from these discrete values
does correspond to a generic (oscillating) type.

\begin{figure}
\centerline{\vbox{\epsfysize=40mm \epsfbox{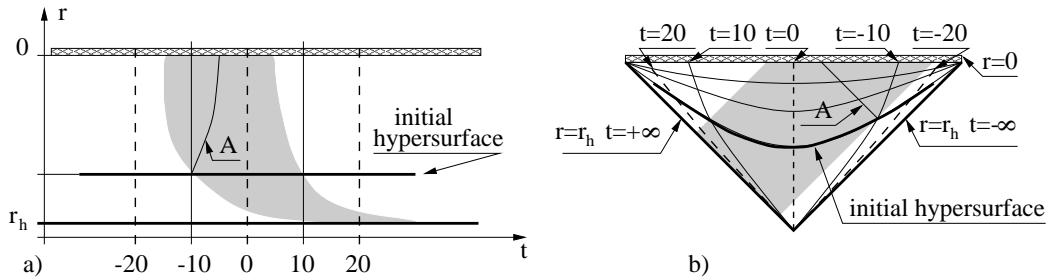}}}
\caption{\large \label{pdiagr} The wide table-like perturbation
imposed at the initial hypersurface came from the horizon
at $t=+\infty$ moves in the negative $t$~-direction
(grey region). In the $r,\, t$ coordinates {\tt a)} the
perturbation almost stops its motion in $t$~-direction and
only goes with time $r$ to the singularity $r=0$. It will never
cross $t=-20$ hypersurface, but it crosses an arbitrary
section $t=\const>0$ in the past with necessity, as it follows
from the conformal diagram, depicted in {\tt b)}. The top of
the ``table'', which corresponds to the cross - section $t=0$ can
reach the singularity $r=0$ before perturbations from the edges
of the initial ``table'' (curve A) will reach it.}
\end{figure}

  It can be expected that small $t$~-dependent perturbations added to
the initial data of these exceptional interiors will produce the
transformation to  some generic type
during the nonlinear evolution towards $r \to 0$.
We have investigated the dynamics of perturbations on various
Schwarzschild  -- type EYM black hole internal solutions and
convinced ourselves that this transformation
really takes place and therefore S-type interiors are occurred
to be unstable.

 Indeed, according to (\ref{classI}), (\ref{classII}) and (\ref{pert}), the
perturbations produce deviations from the background value of the
YM function $f$ in the leading order of series expansions with respect
to $r$ near the event horizon; it is true by definition for class I and
one-wave class of initial perturbations, while the initial perturbations
of class II produce similar deviation of the YM function $f$
at the next step on $r$ as perturbation starts to evolve.

 Thus, only if waves, produced by the initial perturbation,
are suppressed fast enough during their evolution with time $r$ directed
towards $r \to 0$, there is a chance for the exceptional internal solution
to conserve its S-type. However, our numerical studies show that
perturbations are not suppressed during their evolution inward
S-type EYM black hole interior and, as a result, S-type singularity
transforms to the generic oscillating type.

 To illustrate this process it is convenient to investigate
S-type interior, perturbed by the initial table-like ``outgoing''
deviation from the background,
which produces the corresponding wave, propagating with time $r$
from $t=+\infty$ to $t<0$. We integrated
the equation (\ref{constr2}) from the left to the right, the region with
large enough negative $t$ corresponds to the originally unperturbed
S-type solution since the perturbation never can reach this region
[see Fig.\ref{pdiagr}].

Investigating the small $r$~-vicinity of the singularity in
different  spatial $t$~-regions one can answer the questions about the
type of the resulting solution.

We chose the shape of the initial  perturbations  of $f$
to be a Gauss - like curve, broken at the top by large table - like
insertion [see Fig.\ref{fshwinit}]:

\begin{figure}
\centerline{\vbox{\epsfysize=60mm \epsfbox{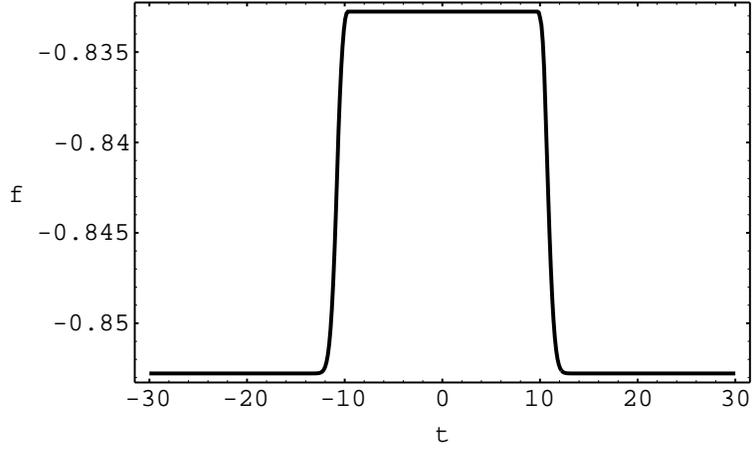}}}
\caption{\large \label{fshwinit}
    Initial $f$ perturbation for the Schwarzschild - type solution.}
\end{figure}

$$
\delta f(t)= \left\{
              \begin{array}{ll}
              K e^{-s(t-t_{01})^2},&t<t_{01},\\
              K,&t_{01}\leq t\leq t_{02},\\
              K e^{-s(t-t_{02})^2},&t>t_{02}.\\
              \end{array}
       \right.
$$

Values of the parameters for the initial data were set equal to:
$$ \begin{array}{rcl}
r_h&=&0.613861419,\\
f_0&=&-0.8478649145,\\
\sigma_0&=&0.289427236,\\
K&=&0.02,\\
t_{01}&=&-10.0,\\
t_{02}&=&10.0.
\end{array}$$
This internal background S-type configuration corresponds to
the asymptotically flat (black hole) solution with $N=1$ (nodes of
YM function) in the external region.
We choose the value of $\phi$  in the way that the outgoing wave goes from
$t=+\infty$ to $t<0$:
\begin{equation}\label{phishw}
\phi(t)= \left\{
              \begin{array}{ll}
              -d_1 \sigma_0 f_1 h/r_h^2 + 2 K s (t-t_{01}) e^{-s(t-t_{01})^2},&t<t_{01},\\
              -d_1 \sigma_0 f_1 h/r_h^2,&t_{01}\leq t\leq t_{02},\\
              -d_1 \sigma_0 f_1 h/r_h^2 + 2 K s (t-t_{02}) e^{-s(t-t_{02})^2},&t>t_{02}.\\
              \end{array}
       \right.
\end{equation}

This shape of the initial perturbations
permits us to investigate three different spatial cross - sections
$t=-20$, $t=0$ and $t=20$ independently,
since the perturbations from the edges of the initial ``table''
(at $t=-10$ and $t=10$) can not reach these points during the evolution
[see Fig.\ref{pdiagr}].

\begin{figure}
\centerline{\vbox{\epsfysize=120mm \epsfbox{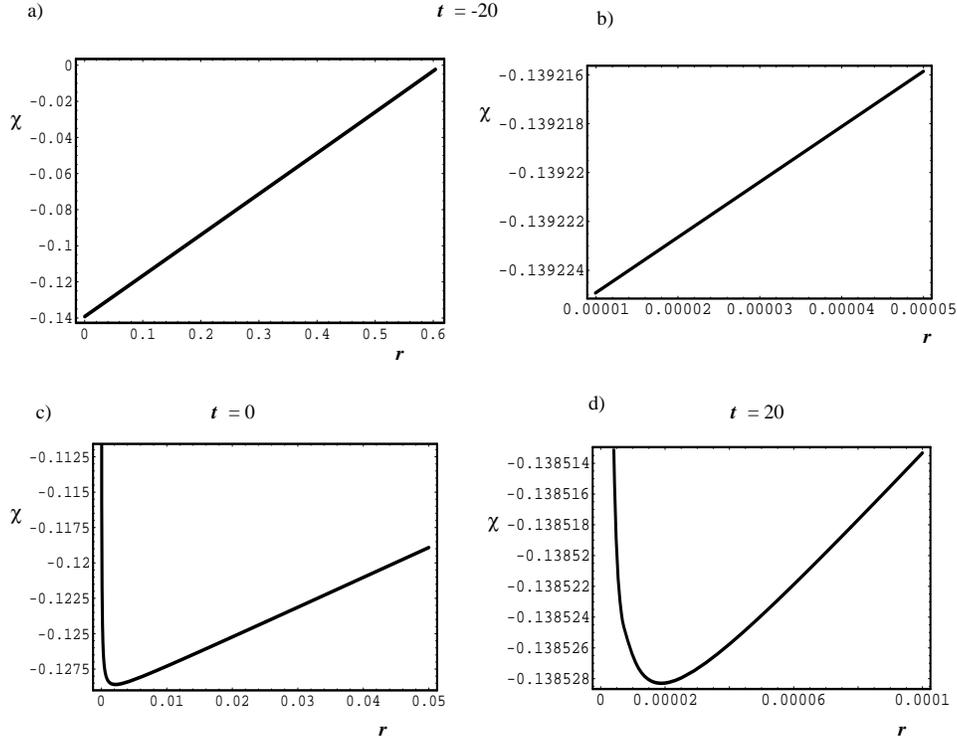}}}
\caption{\large \label{shchi}
  Various cross - sections of $\chi$ for the Schwarzschild  - type
  internal solution. The cross - section $t=-20$ (a) corresponds to the
  non-perturbed background solution: approaching the singularity
  $r=0$, the function $\chi$ goes to the constant (b),
while for any generic (oscillating) internal solution the function
$\chi$ goes to 0 as $r\to 0$. The cross - sections $t=0$ and $t=+20$
(c), (d) demonstrates that the internal solution becomes the generic one.
    }
\end{figure}

The solution in the first spatial cross - section corresponds to the
background Schwarzschild  - type solution. Initial perturbations cannot
reach this region, and we  obtain the typical Schwarzschild - like
behavior [see Fig.\ref{shchi} a) and b)]: the function $\chi$ goes to the constant, 
approaching the singularity $r=0$ [Fig.\ref{shchi} b)].

The generic - like $\chi$ goes to 0 as $r\to 0$. It is easy to see that
the cross - section through the top of the ``table'' at $t=0$
demonstrates just this type of the behavior [Fig.\ref{shchi} c)],
 since the considered perturbation are not suppressed during the
evolution towards $r \to 0$ and the singularity becomes of the
generic (oscillating) type.

Considering
a small vicinity of $r=0$, one can see that the resulting solution
has already the generic type [Fig.\ref{shchi} d)] also in the
third spatial cross - section $t=+20$. The transformation
of S-type to the generic one in this region is caused by the shift
of the apparent horizon position. The pulse of the considered perturbation
shifts the ADM mass of the system and the position of the apparent horizon
$r_h \rightarrow r_h+\delta r_h$, while the relevant value of the
YM function on the apparent horizon $f_0$ remains unperturbed. As
a result, new initial data ($r_h+\delta r_h, f_0$) for the homogeneous system
(\ref{24PRD}) now corresponds to the generic type of the internal solution.

\section{Conclusions}
We have investigated the dynamical evolution of small initial perturbations
in space-time regions which correspond to internal part of
a spherically symmetric black hole in non-Abelian purely
magnetic $SU(2)$ EYM theory.
The obtained results give a strong numerical evidence in favor of
nonlinear stability of the generic (oscillating) type of EYM black
hole interiors while the exceptional Schwarzschild - type interiors
turn out to be unstable and transform to the generic type  as
perturbations are developed towards a singularity.

Now one can expect  that the generic (oscillating) type of the EYM
black hole singularity is stable as well with respect to the
perturbations, penetrating into the internal region from the
exterior through the event horizon. Moreover,
the generic interior solution can pretend to be the final stage of
a spherically symmetric collapse
of the Yang-Mills field after the event horizon is formed.
To check this expectations one should use the null coordinates
and the more precise software tools to attack the problem numerically.
This work is in progress now.

\acknowledgements
\vglue 0.2cm
This work has been supported in part by Russian Foundation for Basic
Research, Grant \mbox{96-02-18126}.

\end{document}